# Comment on "Uncertainty Relation for Photons"

**Zhi-Yong Wang[*], Cai-Dong Xiong, and Qi Qiu**

*School of Optoelectronic Information, University of Electronic Science and Technology of China, Chengdu 610054, CHINA*

In a recent interesting Letter [Phys. Rev. Lett. **108**, 140401 (2012)] I. Bialynicki-Birula and his coauthor have derived the uncertainty relation for the photons in three dimensions. However, some of their arguments are problematical, and this impacts their conclusion.

## 1. The authors of Ref. [1] have confused helicity with spin

Let $\hat{A}^\mu(x) = (\hat{A}^0, \hat{\boldsymbol{A}})$ be a four-dimensional (4D) electromagnetic potential, in the Lorentz gauge condition, it can be expanded as ($k \cdot x = \omega t - \boldsymbol{k} \cdot \boldsymbol{r}$):

$$\hat{A}^\mu(x) = \int \frac{\mathrm{d}^3 k}{\sqrt{2\omega(2\pi)^3}} \sum_{s=0}^{3} \eta^\mu(\boldsymbol{k},s) [\hat{c}(\boldsymbol{k},s) \exp(-\mathrm{i}k \cdot x) + \hat{c}^\dagger(\boldsymbol{k},s) \exp(\mathrm{i}k \cdot x)], \quad (1)$$

where $\eta^\mu(\boldsymbol{k},s)$ ($s = 0,1,2,3$, $\mu = 0,1,2,3$) are four 4D polarization vectors, the four indices of $s=0,1,2,3$ describe four kinds of photons, respectively. One can choose

$$\begin{cases} \eta^\mu(\boldsymbol{k},0) = (1,0,0,0), & \eta^\mu(\boldsymbol{k},1) = (0, \boldsymbol{\varepsilon}(\boldsymbol{k},1)) \\ \eta^\mu(\boldsymbol{k},2) = (0, \boldsymbol{\varepsilon}(\boldsymbol{k},2)), & \eta^\mu(\boldsymbol{k},3) = (0, \boldsymbol{\varepsilon}(\boldsymbol{k},3)) \end{cases}, \quad (2)$$

where $\boldsymbol{\varepsilon}(\boldsymbol{k},i)$ ($i = 1,2,3$) are the 3D linear polarization vectors whose matrix forms are

$$\boldsymbol{\varepsilon}(\boldsymbol{k},1) = \frac{1}{|\boldsymbol{k}|} \begin{pmatrix} \frac{k_1^2 k_3 + k_2^2 |\boldsymbol{k}|}{k_1^2 + k_2^2} \\ \frac{k_1 k_2 (k_3 - |\boldsymbol{k}|)}{k_1^2 + k_2^2} \\ -k_1 \end{pmatrix}, \quad \boldsymbol{\varepsilon}(\boldsymbol{k},2) = \frac{1}{|\boldsymbol{k}|} \begin{pmatrix} \frac{k_1 k_2 (k_3 - |\boldsymbol{k}|)}{k_1^2 + k_2^2} \\ \frac{k_1^2 |\boldsymbol{k}| + k_2^2 k_3}{k_1^2 + k_2^2} \\ -k_2 \end{pmatrix}, \quad \boldsymbol{\varepsilon}(\boldsymbol{k},3) = \frac{\boldsymbol{k}}{|\boldsymbol{k}|} = \frac{1}{|\boldsymbol{k}|} \begin{pmatrix} k_1 \\ k_2 \\ k_3 \end{pmatrix}. \quad (3)$$

One has $\boldsymbol{\varepsilon}(\boldsymbol{k},1) \times \boldsymbol{\varepsilon}(\boldsymbol{k},2) = \boldsymbol{\varepsilon}(\boldsymbol{k},3) = \boldsymbol{k}/|\boldsymbol{k}|$. Obviously, let $\boldsymbol{k} = (0,0,k_3)$ and $k_3 = |\boldsymbol{k}| \geq 0$,

one has $\varepsilon(\boldsymbol{k},1)=(1,0,0)$, $\varepsilon(\boldsymbol{k},2)=(0,1,0)$, and $\varepsilon(\boldsymbol{k},3)=(0,0,1)$, where $\varepsilon(\boldsymbol{k},1)$ and $\varepsilon(\boldsymbol{k},2)$ (perpendicular to $\boldsymbol{k}$) are two transverse polarization vectors, while $\varepsilon(\boldsymbol{k},3)$ (parallel to $\boldsymbol{k}$) is the longitudinal polarization vector, in the 3D space they satisfy the orthonormality and completeness relations (T denotes the matrix transpose, $I_{3\times 3}$ denotes the 3×3 unit matrix):

$$\varepsilon^{\mathrm{T}}(\boldsymbol{k},i)\varepsilon(\boldsymbol{k},j)=\delta_{ij},\quad \sum_i \varepsilon(\boldsymbol{k},i)e^{\mathrm{T}}(\boldsymbol{k},i)=I_{3\times 3},\quad i,j=1,2,3. \qquad (4)$$

The spinor representations of $\varepsilon(\boldsymbol{k},i)$ ($i=1,2,3$) form the circular polarization vectors, i.e.,

$$e_1(\boldsymbol{k})=e_{-1}^*(\boldsymbol{k})=\frac{\varepsilon(\boldsymbol{k},1)+i\varepsilon(\boldsymbol{k},2)}{\sqrt{2}}=\frac{1}{\sqrt{2}|\boldsymbol{k}|}\begin{pmatrix}\frac{k_1 k_3-ik_2|\boldsymbol{k}|}{k_1-ik_2}\\ \frac{k_2 k_3+ik_1|\boldsymbol{k}|}{k_1-ik_2}\\ -(k_1+ik_2)\end{pmatrix},\quad e_0(\boldsymbol{k})=\varepsilon(\boldsymbol{k},3), \qquad (5)$$

where $e_{-1}^*(\boldsymbol{k})$ denotes the complex conjugate of $e_{-1}(\boldsymbol{k})$ (while $e_{-1}^\dagger(\boldsymbol{k})$ denotes the hermitian conjugate of $e_{-1}(\boldsymbol{k})$, and so on). Using Eq. (5) one can prove the orthonormality and completeness relations as follows:

$$e_\lambda^\dagger(\boldsymbol{k})e_{\lambda'}(\boldsymbol{k})=\delta_{\lambda\lambda'},\quad \sum_\lambda e_\lambda(\boldsymbol{k})e_\lambda^\dagger(\boldsymbol{k})=I_{3\times 3},\quad \lambda,\lambda'=\pm 1,0. \qquad (6)$$

In 3D space, the spin matrices of the electromagnetic field are:

$$\tau_1=\begin{pmatrix}0 & 0 & 0\\ 0 & 0 & -i\\ 0 & i & 0\end{pmatrix},\quad \tau_2=\begin{pmatrix}0 & 0 & i\\ 0 & 0 & 0\\ -i & 0 & 0\end{pmatrix},\quad \tau_3=\begin{pmatrix}0 & -i & 0\\ i & 0 & 0\\ 0 & 0 & 0\end{pmatrix}. \qquad (7)$$

They form the spin matrix vector $\boldsymbol{\tau}=(\tau_1,\tau_2,\tau_3)$. Let $\boldsymbol{\tau}\cdot\boldsymbol{k}=\tau_1 k_1+\tau_2 k_2+\tau_3 k_3$, using Eqs. (5) and (7) one can prove that

$$\frac{\boldsymbol{\tau}\cdot\boldsymbol{k}}{|\boldsymbol{k}|}e_\lambda(\boldsymbol{k})=\lambda e_\lambda(\boldsymbol{k}),\quad \lambda=\pm 1,0. \qquad (8)$$

Eq. (8) implies that $e_0(\boldsymbol{k})$ (parallel to $\boldsymbol{k}$) denotes the longitudinal polarization vector,

while $e_{\pm 1}(\boldsymbol{k})$ (perpendicular to $\boldsymbol{k}$) correspond to the right- and left-hand circular polarization vectors, respectively, and $\lambda = \pm 1, 0$ represent the spin projections in the direction of $\boldsymbol{k}$ (i.e., $\lambda = \pm 1, 0$ represent the helicities of photons).

For simplicity let us consider the electromagnetic field in vacuum. In the units of $\hbar = c = 1$, in terms of the electromagnetic field intensities $\hat{\boldsymbol{E}}(\boldsymbol{r},t)$ and $\hat{\boldsymbol{B}}(\boldsymbol{r},t)$ one can define the Riemann-Silberstein vectors $\hat{\boldsymbol{F}}^{(\pm 1)}(\boldsymbol{r},t) = [\hat{\boldsymbol{E}}(\boldsymbol{r},t) \pm i\hat{\boldsymbol{B}}(\boldsymbol{r},t)]/\sqrt{2}$. Substituting Eq. (1) into $\hat{\boldsymbol{E}} = -\nabla \hat{A}^0 - \partial \hat{\boldsymbol{A}}/\partial t$ and $\hat{\boldsymbol{B}} = \nabla \times \hat{\boldsymbol{A}}$ one can prove that (note that $\omega = |\boldsymbol{k}|$)

$$\hat{\boldsymbol{E}} = \int \frac{d^3 k}{\sqrt{2\omega(2\pi)^3}} \sum_{i=1}^{3} \{|\boldsymbol{k}| \varepsilon(\boldsymbol{k},i)[\hat{b}(\boldsymbol{k},i)\exp(-ik \cdot x) + \hat{b}^\dagger(\boldsymbol{k},i)\exp(ik \cdot x)]\}, \quad (9)$$

$$\hat{\boldsymbol{B}} = \int \frac{d^3 k}{\sqrt{2\omega(2\pi)^3}} \sum_{i=1}^{3} \{\boldsymbol{k} \times \varepsilon(\boldsymbol{k},i)[\hat{b}(\boldsymbol{k},i)\exp(-ik \cdot x) + \hat{b}^\dagger(\boldsymbol{k},i)\exp(ik \cdot x)]\}, \quad (10)$$

where

$$\hat{b}(\boldsymbol{k},1) = i\hat{c}(\boldsymbol{k},1), \quad \hat{b}(\boldsymbol{k},2) = i\hat{c}(\boldsymbol{k},2), \quad \hat{b}(\boldsymbol{k},3) = i[\hat{c}(\boldsymbol{k},3) - \hat{c}(\boldsymbol{k},0)]. \quad (11)$$

Using Eq. (3) one can obtain $\boldsymbol{k} \times \varepsilon(\boldsymbol{k},1) = |\boldsymbol{k}|\varepsilon(\boldsymbol{k},2)$, $\boldsymbol{k} \times \varepsilon(\boldsymbol{k},2) = -|\boldsymbol{k}|\varepsilon(\boldsymbol{k},1)$, and $\boldsymbol{k} \times \varepsilon(\boldsymbol{k},3) = (0,0,0)$. Therefore, when the electromagnetic field is described by the 4D electromagnetic potential $\hat{A}^\mu(x)$, there involves four 4D polarization vectors $\eta^\mu(\boldsymbol{k},s)$ ($s = 0, 1, 2, 3$) together describing four kinds of photons; while described by the electromagnetic field intensities $\hat{\boldsymbol{E}}$ and $\hat{\boldsymbol{B}}$, there only involves three 3D polarization vectors $\varepsilon(\boldsymbol{k},i)$ ($i = 1, 2, 3$), and Eq. (11) shows that the $i = 1, 2$ solutions describe two kinds of transverse photons (s=1, 2), while the $i = 3$ photons correspond to the admixture of the longitudinal (s=3) and scalar (s=0) photons. According to QED, only those state vectors (say, $|\Phi\rangle$) are admitted for which the expectation value of the Lorentz gauge condition is satisfied: $\langle\Phi|\partial^\mu \hat{A}_\mu|\Phi\rangle = 0$, which implies that

$$\langle\Phi|\hat{b}(\boldsymbol{k},3)|\Phi\rangle = \mathrm{i}\langle\Phi|[\hat{c}(\boldsymbol{k},3)-\hat{c}(\boldsymbol{k},0)]|\Phi\rangle = 0. \tag{12}$$

Then, we will only take into account the transverse photons. Define

$$\hat{a}_{\pm 1}(\boldsymbol{k}) = \sqrt{\omega/2}[\hat{b}(\boldsymbol{k},1)\mp\mathrm{i}\hat{b}(\boldsymbol{k},2)], \quad \hat{a}_0(\boldsymbol{k}) = \sqrt{\omega/2}\hat{b}(\boldsymbol{k},3), \tag{13}$$

one can prove that $\hat{\boldsymbol{F}}^{(\pm 1)}(\boldsymbol{r},t) = [\hat{\boldsymbol{E}}(\boldsymbol{r},t)\pm\mathrm{i}\hat{\boldsymbol{B}}(\boldsymbol{r},t)]/\sqrt{2}$ are given by

$$\hat{\boldsymbol{F}}^{(1)}(\boldsymbol{r},t) = \int\frac{\mathrm{d}^3 k}{(2\pi)^{3/2}}\boldsymbol{e}_1(\boldsymbol{k})[\hat{a}_1(\boldsymbol{k})\exp(-\mathrm{i}k\cdot x)+\hat{a}^\dagger_{-1}(\boldsymbol{k})\exp(\mathrm{i}k\cdot x)], \tag{14}$$

$$\hat{\boldsymbol{F}}^{(-1)}(\boldsymbol{r},t) = \int\frac{\mathrm{d}^3 k}{(2\pi)^{3/2}}\boldsymbol{e}_{-1}(\boldsymbol{k})[\hat{a}_{-1}(\boldsymbol{k})\exp(-\mathrm{i}k\cdot x)+\hat{a}^\dagger_{1}(\boldsymbol{k})\exp(\mathrm{i}k\cdot x)], \tag{15}$$

where Eq. (14) is equivalent to Eq. (9) of Ref. [1]. The annihilation and creation operators in Eq. (1) satisfy the commutation relations,

$$[\hat{c}(\boldsymbol{k}',s'),\hat{c}^\dagger(\boldsymbol{k},s)] = -g_{ss'}\delta^{(3)}(\boldsymbol{k}-\boldsymbol{k}'), \quad s,s' = 0,1,2,3, \tag{16}$$

where $\delta^{(3)}(\boldsymbol{k}'-\boldsymbol{k}) = \delta(k'_1-k_1)\delta(k'_2-k_2)\delta(k'_3-k_3)$, $g_{ss'} = \mathrm{diag}(1,-1,-1,-1)$. Using Eqs. (11), (13) and (16), one has

$$[\hat{a}_\lambda(\boldsymbol{k}),\hat{a}^\dagger_{\lambda'}(\boldsymbol{k}')] = \omega\delta_{\lambda\lambda'}\delta^{(3)}(\boldsymbol{k}-\boldsymbol{k}'), \quad \lambda,\lambda' = \pm 1, \tag{17}$$

with the others vanishing. In particular, one has $[\hat{a}_0(\boldsymbol{k}),\hat{a}^\dagger_0(\boldsymbol{k}')] = 0$.

Eq. (8) implies that the circular polarization vectors $\boldsymbol{e}_{\pm 1}(\boldsymbol{k})$ are the eienvectors of photonic helicity operator (with the eigenvalues of $\lambda = \pm 1$, respectively), which implies that a photon with the polarization vector $\boldsymbol{e}_\lambda(\boldsymbol{k})$ has the spin projection of $\lambda = \pm 1$ onto the direction of the photon's momentum, and then Eqs. (14) and (15) imply that $\hat{\boldsymbol{F}}^{(1)}(\boldsymbol{r},t)$ and $\hat{\boldsymbol{F}}^{(-1)}(\boldsymbol{r},t)$ describe the transverse photons with the helicities of $\lambda = \pm 1$, respectively. On the other hand, taking $\hat{\boldsymbol{F}}^{(1)}(\boldsymbol{r},t)$ for example, if its positive-frequency part has the momentum of $\boldsymbol{k}\in(-\infty,+\infty)$, then its negative-frequency part has the momentum of $-\boldsymbol{k}$. As a result, if the state vector of $\hat{a}^\dagger_1(\boldsymbol{k})|0\rangle$ has the spin of "up", then the one of $\hat{a}^\dagger_{-1}(\boldsymbol{k})|0\rangle$

has the spin of "down", but both of them have positive helicity (because $\hat{\boldsymbol{F}}^{(1)}(\boldsymbol{r},t)$ has positive helicity). That is, both the annihilation and creation operators in Eq. (9) of Ref. [1] have positive helicity, while their spins are respectively "up" and "down". Therefore, the authors of Ref. [1] have confused helicity with spin, and the statements before Eq. (9) in Ref. [1] are not appropriate.

BTW, in terms of our circular polarization vector $\boldsymbol{e}_1(\boldsymbol{k})$, one can express the normalized vector $\boldsymbol{e}(\boldsymbol{k})$ given by Eq. (10) in Ref. [1] as

$$\boldsymbol{e}(\boldsymbol{k}) = -\frac{k_1 - ik_2}{k_1^2 + k_2^2}\boldsymbol{e}_1(\boldsymbol{k}) = \exp(i\theta)\boldsymbol{e}_1(\boldsymbol{k}), \quad \theta = -\arctan(k_2/k_1), \qquad (18)$$

Then the normalized vector $\boldsymbol{e}(\boldsymbol{k})$ is also the eigenvector of Eq. (8) with the eigenvalue $\lambda = 1$, i.e., it is also the right-hand circular polarization vector. In fact, rotating $\boldsymbol{e}_{\pm 1}(\boldsymbol{k})$ round the wave number vector $\boldsymbol{k}$, one can obtain another right- and left-hand circular polarization vectors, respectively. The polarization vectors $\boldsymbol{e}_{\pm 1}(\boldsymbol{k})$ have the following properties ($\lambda = \pm 1$):

$$\boldsymbol{e}_\lambda^\dagger(-\boldsymbol{k})\boldsymbol{e}_\lambda(\boldsymbol{k}) = 0, \quad [\frac{\partial}{\partial \boldsymbol{k}}\boldsymbol{e}_\lambda^\dagger(-\boldsymbol{k})]\boldsymbol{e}_\lambda(\boldsymbol{k}) = (0,0,0), \quad [\frac{\partial^2}{\partial \boldsymbol{k}^2}\boldsymbol{e}_\lambda^\dagger(-\boldsymbol{k})]\boldsymbol{e}_\lambda(\boldsymbol{k}) = 0, \qquad (20)$$

$$[\frac{\partial}{\partial \boldsymbol{k}}\boldsymbol{e}_\lambda^\dagger(\boldsymbol{k})]\boldsymbol{e}_\lambda(\boldsymbol{k}) = \frac{-i\lambda}{|\boldsymbol{k}|(|\boldsymbol{k}|+k_3)}(-k_2,k_1,0), \quad \frac{\partial}{\partial \boldsymbol{k}}\cdot\{[\frac{\partial}{\partial \boldsymbol{k}}\boldsymbol{e}_\lambda^\dagger(\boldsymbol{k})]\boldsymbol{e}_\lambda(\boldsymbol{k})\} = 0, \qquad (21)$$

$$[\frac{\partial^2}{\partial \boldsymbol{k}^2}\boldsymbol{e}_\lambda^\dagger(\boldsymbol{k})]\boldsymbol{e}_\lambda(\boldsymbol{k}) = \frac{-2}{|\boldsymbol{k}|(|\boldsymbol{k}|+k_3)}. \qquad (22)$$

Note that $\partial^2/\partial \boldsymbol{k}^2 = (\partial/\partial \boldsymbol{k})\cdot(\partial/\partial \boldsymbol{k}) = \partial^2/\partial k_1^2 + \partial^2/\partial k_2^2 + \partial^2/\partial k_3^2$, and so on. The vector $\boldsymbol{e}_\lambda(\boldsymbol{k})$ is expressed in matrix forms, such that the scalar product of $\boldsymbol{e}_\lambda(\boldsymbol{k})$ with itself is $\boldsymbol{e}_\lambda^\dagger(\boldsymbol{k})\boldsymbol{e}_\lambda(\boldsymbol{k})$, for example. However, the vector such as $[\partial \boldsymbol{e}_\lambda^\dagger(\boldsymbol{k})/\partial \boldsymbol{k}]\boldsymbol{e}_\lambda(\boldsymbol{k})$ is expressed in the usual form, because its vectorial property comes from $\partial/\partial \boldsymbol{k}$.

**2. Eq. (15) (and then Eq. (27)) in Ref. [1] is wrong**

To show this, similar to Ref. [1], we replace all field operators with the corresponding classical fields, and rewrite Eqs. (14) and (15) as ($\hbar = c = 1$)

$$\boldsymbol{F}^{(\lambda)}(\boldsymbol{r},t) = \int \frac{\mathrm{d}^3 k}{(2\pi)^{3/2}} \boldsymbol{e}_\lambda(\boldsymbol{k})[a_\lambda(\boldsymbol{k})\exp(-\mathrm{i}k\cdot x) + a_{-\lambda}^*(\boldsymbol{k})\exp(\mathrm{i}k\cdot x)]. \tag{23}$$

Note that our $a_{\pm 1}(\boldsymbol{k})$ are identical with $f_\pm(\boldsymbol{k})$ in Ref. [1], while our $\hat{a}_{\pm 1}(\boldsymbol{k})$ are identical with $a_\pm(\boldsymbol{k})$ in Ref. [1]. To calculate an energy moment, one should simultaneously take into account two kinds of transverse photons with the helicities of $\lambda = \pm 1$, which also lies in the fact that, the field quantities of $\boldsymbol{E}(\boldsymbol{r},t)$ and $\boldsymbol{B}(\boldsymbol{r},t)$ together are equivalent to the ones of $\boldsymbol{F}^{(1)}(\boldsymbol{r},t)$ and $\boldsymbol{F}^{(-1)}(\boldsymbol{r},t)$ together, rather than to $\boldsymbol{F}^{(1)}(\boldsymbol{r},t)$ only. Then, in spite of $|\boldsymbol{F}^{(1)}| = |\boldsymbol{F}^{(-1)}|$, conceptually, the first and second moments of the classical energy density should be

$$M_1 = \frac{1}{2}\int \mathrm{d}^3 r\, \boldsymbol{r}(|\boldsymbol{F}^{(1)}|^2 + |\boldsymbol{F}^{(-1)}|^2), M_2 = \frac{1}{2}\int \mathrm{d}^3 r\, \boldsymbol{r}^2[|\boldsymbol{F}^{(1)}|^2 + |\boldsymbol{F}^{(-1)}|^2]. \tag{24}$$

For the moment one has $a_\lambda(\boldsymbol{k})a_{\lambda'}^*(\boldsymbol{k}') = a_{\lambda'}^*(\boldsymbol{k}')a_\lambda(\boldsymbol{k})$ ($\lambda, \lambda' = \pm 1$). For $t = 0$, one can prove that (see **Appendix A**)

$$M_1 = -\mathrm{i}\int \mathrm{d}^3 k[a_1(\boldsymbol{k})\boldsymbol{D}_{\boldsymbol{k}} a_1^*(\boldsymbol{k}) - a_{-1}^*(\boldsymbol{k})\boldsymbol{D}_{\boldsymbol{k}} a_{-1}(\boldsymbol{k})], \tag{25}$$

$$M_2 = \int \mathrm{d}^3 k\{\frac{1}{|\boldsymbol{k}|^2}[|a_1(\boldsymbol{k})|^2 + |a_{-1}(\boldsymbol{k})|^2] - [a_1(\boldsymbol{k})\boldsymbol{D}_{\boldsymbol{k}}^2 a_1^*(\boldsymbol{k}) + a_{-1}^*(\boldsymbol{k})\boldsymbol{D}_{\boldsymbol{k}}^2 a_{-1}(\boldsymbol{k})]\}, \tag{26}$$

where $\boldsymbol{D}_{\boldsymbol{k}} = \partial/\partial \boldsymbol{k} - \mathrm{i}\boldsymbol{\varLambda}$ and $\boldsymbol{\varLambda} = (-k_2, k_1, 0)/|\boldsymbol{k}|(|\boldsymbol{k}| + k_3)$. Our Eq. (26) is different from Eq. (15) in Ref. [1], such that in Ref. [1] the conclusion based on Eqs. (15) and (27) are questionable. Using the facts that $M_1$ is a real number and $M_2$ is nonegative definite, one can obtain some relations.

### 3. The definition Eq. (8) in Ref. [1] is not reasonable

The classical energy is

$$M_0 = (1/2)\int d^3r[|\boldsymbol{F}^{(1)}|^2 + |\boldsymbol{F}^{(-1)}|^2] = \int d^3k[|a_1(\boldsymbol{k})|^2 + |a_{-1}(\boldsymbol{k})|^2]. \tag{27}$$

Let us denote

$$\langle \boldsymbol{r}^n \rangle = \frac{M_n}{M_0} = \frac{(1/2)\int d^3r \boldsymbol{r}^n (|\boldsymbol{F}^{(+)}|^2 + |\boldsymbol{F}^{(-)}|^2)}{(1/2)\int d^3r (|\boldsymbol{F}^{(+)}|^2 + |\boldsymbol{F}^{(-)}|^2)}, \quad n=1,2. \tag{28}$$

Obviously, $\langle \boldsymbol{r}^n \rangle = M_n/M_0$ have the dimension of [length]$^n$. One can define

$$(\Delta r)^2 = \langle \boldsymbol{r}^2 \rangle - \langle \boldsymbol{r} \rangle^2. \tag{29}$$

We do not think the definition Eq. (8) in Ref. [1] is reasonable. Instead, we think that $\Delta r$ should be defined via Eq. (29).

To provide a heuristic insight, let us consider a single-mode field with the frequency of $\omega = |\boldsymbol{k}|$ and let $a_{-1}(\boldsymbol{k}) = 0$, $|a_1(\boldsymbol{k})|^2 = 1$. For the moment one has $\partial a_1^*(\boldsymbol{k})/\partial \boldsymbol{k} = 0$, $M_0 = 1$, $\langle \boldsymbol{r} \rangle = M_1 = -\boldsymbol{\Lambda}$, $\langle \boldsymbol{r}^2 \rangle = M_2 = -L$, and then $\Delta r = \sqrt{-L - |\boldsymbol{\Lambda}|^2} = 1/|\boldsymbol{k}|$.

## 4. Conclusion

The statements before Eq. (9) in Ref. [1] have confused the concept of helicity with that of spin; Eq. (15) in Ref. [1] should be replaced with our Eq. (26). As a result, in Ref. [1] the conclusions related to (15) and Eq. (27), etc., have not been proven. Moreover, conceptually, we do not think the definition Eq. (8) in Ref. [1] is reasonable. Instead, we think that $\Delta r$ should be defined via Eq. (29).

---

*E-mail:  zywang@uestc.edu.cn

[1] I. Bialynicki-Birula and Z. Bialynicka-Birula, Phys. Rev. Lett. **108**, 140401 (2012).

## Appendix A Proof of Eq. (26)

Let us calculate the second moment of the classical energy density

$$M_2 = \frac{1}{2}\int d^3r r^2 (|\mathbf{F}^{(1)}|^2 + |\mathbf{F}^{(-1)}|^2) = \int d^3r r^2 |\mathbf{F}^{(1)}|^2, \tag{a1}$$

where ($k \cdot x = \omega t - \mathbf{k} \cdot \mathbf{r}$)

$$\mathbf{F}^{(1)}(\mathbf{r},t) = \int \frac{d^3k}{(2\pi)^{3/2}} \mathbf{e}_1(\mathbf{k})[a_1(\mathbf{k})\exp(-ik \cdot x) + a_{-1}^*(\mathbf{k})\exp(ik \cdot x)]. \tag{a2}$$

Substituting (a2) into (a1), one has (note that the vectors $\mathbf{e}_1(\mathbf{k})$ is expressed in matrix form, such that the scalar product of $\mathbf{e}_1(\mathbf{k})$ with itself is $\mathbf{e}_1^\dagger(\mathbf{k})\mathbf{e}_1(\mathbf{k})$):

$$M_2 = \int d^3r r^2 \{\int \frac{d^3k'}{(2\pi)^{3/2}} \mathbf{e}_1^\dagger(\mathbf{k}')[a_1^*(\mathbf{k}')\exp(ik' \cdot x) + a_{-1}(\mathbf{k}')\exp(-ik' \cdot x)]$$

$$\int \frac{d^3k}{(2\pi)^{3/2}} \mathbf{e}_1(\mathbf{k})[a_1(\mathbf{k})\exp(-ik \cdot x) + a_{-1}^*(\mathbf{k})\exp(ik \cdot x)]\}$$

$$= \int d^3k \int d^3k' \int r^2 \frac{d^3r}{(2\pi)^3}$$

$$\{\mathbf{e}_1^\dagger(\mathbf{k}')a_1^*(\mathbf{k}')\mathbf{e}_1(\mathbf{k})a_1(\mathbf{k})\exp(ik' \cdot x)\exp(-ik \cdot x)$$
$$+\mathbf{e}_1^\dagger(\mathbf{k}')a_{-1}(\mathbf{k}')\mathbf{e}_1(\mathbf{k})a_{-1}^*(\mathbf{k})\exp(-ik' \cdot x)\exp(ik \cdot x)$$
$$+\mathbf{e}_1^\dagger(\mathbf{k}')a_1^*(\mathbf{k}')\mathbf{e}_1(\mathbf{k})a_{-1}^*(\mathbf{k})\exp(ik' \cdot x)\exp(ik \cdot x)$$
$$+\mathbf{e}_1^\dagger(\mathbf{k}')a_{-1}(\mathbf{k}')\mathbf{e}_1(\mathbf{k})a_1(\mathbf{k})\exp(-ik' \cdot x)\exp(-ik \cdot x)\}$$

$$= -\int d^3k \int d^3k' \int \frac{d^3r}{(2\pi)^3}$$

$$\{\mathbf{e}_1^\dagger(\mathbf{k}')a_1^*(\mathbf{k}')\mathbf{e}_1(\mathbf{k})a_1(\mathbf{k})\frac{\partial^2}{\partial \mathbf{k}'^2}\exp(ik' \cdot x)\exp(-ik \cdot x)$$

$$+\mathbf{e}_1^\dagger(\mathbf{k}')a_{-1}(\mathbf{k}')\mathbf{e}_1(\mathbf{k})a_{-1}^*(\mathbf{k})\frac{\partial^2}{\partial \mathbf{k}'^2}\exp(-ik' \cdot x)\exp(ik \cdot x)$$

$$+\mathbf{e}_1^\dagger(\mathbf{k}')a_1^*(\mathbf{k}')\mathbf{e}_1(\mathbf{k})a_{-1}^*(\mathbf{k})\frac{\partial^2}{\partial \mathbf{k}'^2}\exp(ik' \cdot x)\exp(ik \cdot x) \tag{a3}$$

$$+\mathbf{e}_1^\dagger(\mathbf{k}')a_{-1}(\mathbf{k}')\mathbf{e}_1(\mathbf{k})a_1(\mathbf{k})\frac{\partial^2}{\partial \mathbf{k}'^2}\exp(-ik' \cdot x)\exp(-ik \cdot x)\}$$

Using

$$\int \frac{d^3r}{(2\pi)^3}\exp[\pm i(\mathbf{k}'-\mathbf{k})\cdot\mathbf{r}] = \delta^{(3)}(\mathbf{k}'-\mathbf{k}), \int \frac{d^3r}{(2\pi)^3}\exp[\pm i(\mathbf{k}'+\mathbf{k})\cdot\mathbf{r}] = \delta^{(3)}(\mathbf{k}'+\mathbf{k}), \tag{a4}$$

one has

$$M_2 = -\int d^3k \int d^3k'$$

$$\{[e_1^\dagger(k')a_1^*(k')\exp(i\omega't)]e_1(k)a_1(k)\exp(-i\omega t)\frac{\partial^2}{\partial k'^2}\delta^{(3)}(k'-k)$$

$$+[e_1^\dagger(k')a_{-1}(k')\exp(-i\omega't)]e_1(k)a_{-1}^*(k)\exp(i\omega t)\frac{\partial^2}{\partial k'^2}\delta^{(3)}(k'-k) \quad , \qquad (a5)$$

$$+[e_1^\dagger(k')a_1^*(k')\exp(i\omega't)]e_1(k)a_{-1}^*(k)\exp(i\omega t)\frac{\partial^2}{\partial k'^2}\delta^{(3)}(k'+k)$$

$$+[e_1^\dagger(k')a_{-1}(k')\exp(-i\omega't)]e_1(k)a_1(k)\exp(-i\omega t)\frac{\partial^2}{\partial k'^2}\delta^{(3)}(k'+k)\}$$

Using $\partial^2/\partial k'^2 = \partial^2/\partial k_1'^2 + \partial^2/\partial k_2'^2 + \partial^2/\partial k_3'^2$, $\delta^{(3)}(k'-k) = \delta(k_1'-k_1)\delta(k_2'-k_2)\delta(k_3'-k_3)$, and

$$f(x')[\frac{\partial^n}{\partial x'^n}\delta(x'-x)] = (-1)^n \delta(x'-x)[\frac{\partial^n}{\partial x'^n}f(x')], \qquad (a6)$$

one has

$$M_2 = -\int d^3k \int d^3k'$$

$$\{[\frac{\partial^2}{\partial k'^2}e_1^\dagger(k')a_1^*(k')\exp(i\omega't)]e_1(k)a_1(k)\exp(-i\omega t)\delta^{(3)}(k'-k)$$

$$+[\frac{\partial^2}{\partial k'^2}e_1^\dagger(k')a_{-1}(k')\exp(-i\omega't)]e_1(k)a_{-1}^*(k)\exp(i\omega t)\delta^{(3)}(k'-k)$$

$$+[\frac{\partial^2}{\partial k'^2}e_1^\dagger(k')a_1^*(k')\exp(i\omega't)]e_1(k)a_{-1}^*(k)\exp(i\omega t)\delta^{(3)}(k'+k)$$

$$+[\frac{\partial^2}{\partial k'^2}e_1^\dagger(k')a_{-1}(k')\exp(-i\omega't)]e_1(k)a_1(k)\exp(-i\omega t)\delta^{(3)}(k'+k)\},$$

$$= -\int d^3k$$

$$\{[\frac{\partial^2}{\partial k^2}e_1^\dagger(k)a_1^*(k)\exp(i\omega t)]e_1(k)a_1(k)\exp(-i\omega t)$$

$$+[\frac{\partial^2}{\partial k^2}e_1^\dagger(k)a_{-1}(k)\exp(-i\omega t)]e_1(k)a_{-1}^*(k)\exp(i\omega t)$$

$$+[\frac{\partial^2}{\partial k^2}e_1^\dagger(-k)a_1^*(-k)\exp(i\omega t)]e_1(k)a_{-1}^*(k)\exp(i\omega t)$$

$$+[\frac{\partial^2}{\partial k^2}e_1^\dagger(-k)a_{-1}(-k)\exp(-i\omega t)]e_1(k)a_1(k)\exp(-i\omega t)\} \qquad . \quad (a7)$$

$$= -\int d^3k(B_1 + B_2 + B_3 + B_4)$$

where

$$B_1 = [\frac{\partial^2}{\partial k^2} e_1^\dagger(k) a_1^*(k) \exp(i\omega t)] e_1(k) a_1(k) \exp(-i\omega t), \tag{a8}$$

$$B_2 = [\frac{\partial^2}{\partial k^2} e_1^\dagger(k) a_{-1}(k) \exp(-i\omega t)] e_1(k) a_{-1}^*(k) \exp(i\omega t), \tag{a9}$$

$$B_3 = [\frac{\partial^2}{\partial k^2} e_1^\dagger(-k) a_1^*(-k) \exp(i\omega t)] e_1(k) a_{-1}^*(k) \exp(i\omega t), \tag{a10}$$

$$B_4 = [\frac{\partial^2}{\partial k^2} e_1^\dagger(-k) a_{-1}(-k) \exp(-i\omega t)] e_1(k) a_1(k) \exp(-i\omega t). \tag{a11}$$

One can prove that

$$e_1^\dagger(-k) e_1(k) = 0, \quad [\frac{\partial}{\partial k} e_1^\dagger(-k)] e_1(k) = (0,0,0), \quad [\frac{\partial^2}{\partial k^2} e_1^\dagger(-k)] e_1(k) = 0, \tag{a12}$$

and then one has $B_3 = B_4 = 0$, it follows that

$$M_2 = -\int d^3 k$$
$$\{[\frac{\partial^2}{\partial k^2} e_1^\dagger(k) a_1^*(k) \exp(i\omega t)] e_1(k) a_1(k) \exp(-i\omega t) \tag{a13}$$
$$+ [\frac{\partial^2}{\partial k^2} e_1^\dagger(k) a_{-1}(k) \exp(-i\omega t)] e_1(k) a_{-1}^*(k) \exp(i\omega t)\}$$

Here the vectors $e_1(k)$ is expressed in matrix form, such that the scalar product of $e_1(k)$ with itself is $e_1^\dagger(k) e_1(k)$. However, the vector such as $[\partial e_1^\dagger(k)/\partial k] e_1(k)$ is expressed in the usual form, because its vectorial property comes from $\partial/\partial k$. Let $t = 0$, (a13) becomes

$$M_2 = -\int d^3 k$$
$$\{[\frac{\partial^2}{\partial k^2} e_1^\dagger(k) a_1^*(k)] e_1(k) a_1(k) + [\frac{\partial^2}{\partial k^2} e_1^\dagger(k) a_{-1}(k)] e_1(k) a_{-1}^*(k)\}, \tag{a14}$$

where

$$\frac{\partial^2}{\partial k^2} e_1^\dagger(k) a_1^*(k) = [\frac{\partial^2}{\partial k^2} e_1^\dagger(k)] a_1^*(k) + 2[\frac{\partial}{\partial k} e_1^\dagger(k)] \cdot [\frac{\partial}{\partial k} a_1^*(k)] + e_1^\dagger(k) \frac{\partial^2}{\partial k^2} a_1^*(k), \tag{a15}$$

$$\frac{\partial^2}{\partial k^2} e_1^\dagger(k) a_{-1}(k) = [\frac{\partial^2}{\partial k^2} e_1^\dagger(k)] a_{-1}(k) + 2[\frac{\partial}{\partial k} e_1^\dagger(k)] \cdot [\frac{\partial}{\partial k} a_{-1}(k)] + e_1^\dagger(k) \frac{\partial^2}{\partial k^2} a_{-1}(k). \tag{a16}$$

Substituting (a15) and (a16) into (a14), and define

$$\Lambda = \mathrm{i}[\frac{\partial}{\partial \boldsymbol{k}} e_1^\dagger(\boldsymbol{k})]e_1(\boldsymbol{k}) = \frac{1}{|\boldsymbol{k}|(|\boldsymbol{k}|+k_3)}(-k_2, k_1, 0), \quad L = [\frac{\partial^2}{\partial \boldsymbol{k}^2} e_1^\dagger(\boldsymbol{k})]e_1(\boldsymbol{k}) = \frac{-2}{|\boldsymbol{k}|(|\boldsymbol{k}|+k_3)}, \quad (a17)$$

one has (note that $e_1^\dagger(\boldsymbol{k})e_1(\boldsymbol{k}) = 1$)

$$\begin{aligned} M_2 = -\int \mathrm{d}^3 k \{ & L[a_1^*(\boldsymbol{k})a_1(\boldsymbol{k}) + a_{-1}(\boldsymbol{k})a_{-1}^*(\boldsymbol{k})] \\ & -2\mathrm{i}\Lambda \cdot [\frac{\partial}{\partial \boldsymbol{k}} a_1^*(\boldsymbol{k})]a_1(\boldsymbol{k}) - 2\mathrm{i}\Lambda \cdot [\frac{\partial}{\partial \boldsymbol{k}} a_{-1}(\boldsymbol{k})]a_{-1}^*(\boldsymbol{k}), \\ & +[\frac{\partial^2}{\partial \boldsymbol{k}^2} a_1^*(\boldsymbol{k})]a_1(\boldsymbol{k}) + [\frac{\partial^2}{\partial \boldsymbol{k}^2} a_{-1}(\boldsymbol{k})]a_{-1}^*(\boldsymbol{k}) \} \end{aligned} \quad (a18)$$

Let $\boldsymbol{D}_k = \partial/\partial \boldsymbol{k} - \mathrm{i}\Lambda$, using $\frac{\partial}{\partial \boldsymbol{k}} \cdot \Lambda = \partial \Lambda_1/\partial k_1 + \partial \Lambda_2/\partial k_2 + \partial \Lambda_3/\partial k_3 = 0$, one has

$$a_1 \boldsymbol{D}_k^2 a_1^* = a_1(\frac{\partial}{\partial \boldsymbol{k}} - \mathrm{i}\Lambda) \cdot (\frac{\partial}{\partial \boldsymbol{k}} - \mathrm{i}\Lambda)a_1^* = a_1 \frac{\partial^2 a_1^*}{\partial \boldsymbol{k}^2} - 2\mathrm{i}\Lambda \cdot (\frac{\partial}{\partial \boldsymbol{k}} a_1^*)a_1 - |\Lambda|^2 |a_1|^2, \quad (a19)$$

$$a_{-1}^* \boldsymbol{D}_{k,-1}^2 a = a_{-1}^*(\frac{\partial}{\partial \boldsymbol{k}} - \mathrm{i}\Lambda) \cdot (\frac{\partial}{\partial \boldsymbol{k}} - \mathrm{i}\Lambda)a_{-1} = a_{-1}^* \frac{\partial^2 a_{-1}}{\partial \boldsymbol{k}^2} - 2\mathrm{i}\Lambda \cdot (\frac{\partial}{\partial \boldsymbol{k}} a_{-1})a_{-1}^* - |\Lambda|^2 |a_{-1}|^2. \quad (a20)$$

Using (a19) and (a20), and consider that $L + |\Lambda|^2 = -1/|\boldsymbol{k}|^2$ one can obtain Eq. (26), i.e.,

$$\begin{aligned} M_2 &= \frac{1}{2} \int \mathrm{d}^3 r \, r^2 (|\boldsymbol{F}^{(1)}|^2 + |\boldsymbol{F}^{(-1)}|^2) = \int \mathrm{d}^3 r \, r^2 |\boldsymbol{F}^{(1)}|^2 \\ &= \int \mathrm{d}^3 k \{ \frac{1}{|\boldsymbol{k}|^2}[|a_1(\boldsymbol{k})|^2 + |a_{-1}(\boldsymbol{k})|^2] - [a_1(\boldsymbol{k})\boldsymbol{D}_k^2 a_1^*(\boldsymbol{k}) + a_{-1}^*(\boldsymbol{k})\boldsymbol{D}_k^2 a_{-1}(\boldsymbol{k})] \} \end{aligned} \quad (a21)$$

BTW, likewise, one can prove that

$$\begin{aligned} M_1 &= \frac{1}{2} \int \mathrm{d}^3 r \, \boldsymbol{r}(|\boldsymbol{F}^{(1)}|^2 + |\boldsymbol{F}^{(-1)}|^2) = \int \mathrm{d}^3 r \, \boldsymbol{r} |\boldsymbol{F}^{(1)}|^2 \\ &= -\mathrm{i} \int \mathrm{d}^3 k [a_1(\boldsymbol{k})\boldsymbol{D}_k a_1^*(\boldsymbol{k}) - a_{-1}^*(\boldsymbol{k})\boldsymbol{D}_k a_{-1}(\boldsymbol{k})] \end{aligned}, \quad (a22)$$

$$\begin{aligned} M_0 &= \frac{1}{2} \int \mathrm{d}^3 r [|\boldsymbol{F}^{(1)}|^2 + |\boldsymbol{F}^{(-1)}|^2] = \int \mathrm{d}^3 r |\boldsymbol{F}^{(1)}|^2 \\ &= \int \mathrm{d}^3 k [|a_1(\boldsymbol{k})|^2 + |a_{-1}(\boldsymbol{k})|^2] \end{aligned}. \quad (a23)$$